\documentclass[11pt]{article}
\usepackage{epsfig}
\usepackage[usenames]{color}
\usepackage{graphicx}
\usepackage{amsmath}
\def\O{{\cal O}}
\def\la{\langle}\def\ra{\rangle}
\def\be{\begin{eqnarray}}\def\ee{\end{eqnarray}}
\def\lsim{\mathrel{\rlap{\lower3pt\hbox{\hskip1pt$\sim$}}
     \raise1pt\hbox{$<$}}} 
\def\gsim{\mathrel{\rlap{\lower3pt\hbox{\hskip1pt$\sim$}}
     \raise1pt\hbox{$>$}}} 
\def\le{ \begin{array}{ll}}\def\re{\end{array}}

\def\lear{ \left( \begin{array}{cc}}\def\rear{\end{array} \right)}

\def\le{ \left( \begin{array}{cc}}\def\re{\end{array} \right)}

\def\bi{\bibitem}

\def\O{\cal O}

\newcommand{\Rmnum}[1]{\expandafter\@slowromancap\romannumeral #1@}

\renewcommand{\thesection}{\arabic{section}}

\renewcommand{\theequation}{\arabic{equation}}

\def\lsim{\mathrel{\rlap{\lower3pt\hbox{\hskip1pt$\sim$}}
     \raise1pt\hbox{$<$}}} 
\def\gsim{\mathrel{\rlap{\lower3pt\hbox{\hskip1pt$\sim$}}
     \raise1pt\hbox{$>$}}} 

\def\la{\langle}
\def\ra{\rangle}
\def\bi{\bibitem}

\begin{document}

\centerline{\Large\bf Kaon Condensation in Baryonic Fermi Liquid at High Density}


\vskip 0.5cm
\begin{center}
{ Won-Gi Paeng\footnote{\sf e-mail: wgpaeng@ibs.re.kr}
}\\
{\em Rare Isotope Science Project, Institute for Basic Science, Daejeon 305-811,  Korea}

{ Mannque Rho\footnote{\sf e-mail: mannque.rho@cea.fr }
}\\
{\em Institut de Physique Th\'eorique, CEA Saclay, 91191 Gif-sur-Yvette c\'edex, France }

\vskip 0.5cm
{\ (\today)}
\end{center}
\vskip 1cm
\centerline{\large\bf ABSTRACT}
We formulate kaon condensation in dense baryonic matter with anti-kaons fluctuating from the Fermi-liquid fixed point. This entails that in the Wilsonian RG approach, the decimation is effectuated in the baryonic sector to the Fermi surface while in the meson sector to the origin. In writing the kaon-baryon (KN) coupling, we will take a generalized hidden local symmetry Lagrangian for the meson sector endowed with a ``mended symmetry" that has the unbroken symmetry limit at high density in which the Goldstone $\pi$, scalar $s$,  and vectors $\rho$ (and $\omega$) and $a_1$ become massless. The vector mesons $\rho$ (and $\omega$) and $a_1$ can be identified as emergent (hidden) local gauge fields and the scalar $s$ as the dilaton field of the spontaneously broken scale invariance at chiral restoration. In matter-free space, when  the vector mesons and the scalar meson -- whose masses are much greater than that of the pion -- are integrated out, then the resulting KN coupling Lagrangian consists of the leading chiral order ($O(p^1)$) Weinberg-Tomozawa term and the next chiral order ($O(p^2)$) $\Sigma_{KN}$ term. In addressing kaon condensation in dense nuclear matter in chiral perturbation theory (ChPT), one makes an expansion in the ``small" Fermi momentum $k_F$. We argue that in the Wilsonian RG formalism with the Fermi-liquid fixed point, the expansion is on the contrary in $1/k_F$ with the ``large" Fermi momentum $k_F$. The kaon-quasinucleon interaction resulting from integrating out the massive mesons consists of a ``relevant" term from the scalar exchange (analog to the $\Sigma_{KN}$ term) and an ``irrelevant" term from the vector-meson exchange (analog to the Weinberg-Tomozawa term). It is found that the critical density predicted by the latter approach, controlled by the relevant term with the irrelevant term suppressed, is three times less than that predicted by chiral perturbation theory. This would make kaon condensation take place at a much lower density than previously estimated in chiral perturbation theory.

\section{Introduction}
Negatively charged kaons ($K^-$ that will be referred to simply as $K$ or kaon) are predicted in simple chiral Lagrangians to bose-condense in s-wave in dense baryonic matter at some high density~\cite{kaplan-nelson}.  Within the regime of strong interactions such as in relativistic heavy-ion collisions, the critical density for the process -- that we will refer to as kaon condensation -- is likely to be located at $n_K\gsim 6n_0$ where $n_0$ is the ordinary nuclear matter density $n_0\approx 0.16$ fm$^{-3}$. This process may be difficult to access in the laboratories for various reasons, among which the high temperature expected, given that the temperature produced would melt the kaon condensate. It would also be compounded with other strange degrees of freedom such as strange quark matter. In compact-star matter in weak and chemical equilibrium, mean-field applications of chiral Lagrangians  typically predict the critical density considerably lower, $n_K\sim 3n_0$~\cite{BTKR,BLR}. As reviewed in \cite{BLR}, this critical density is arrived at by three different starting points, one from the vacuum ($n=0$), the second from near nuclear matter density ($n\sim n_0$) and third from the vector manifestation fixed point $n_c$ at which the $\rho$-meson mass vanishes (in the chiral limit) and chiral symmetry gets restored. One common element in these three treatments, stressed in \cite{BLR}, is the underlying symmetry that governs the strong interactions, namely,  hidden gauge symmetry implementing chiral symmetry of QCD appropriate at the kinematic regime of density. In this way, both the boson and the fermion involved are treated in one framework as we will elaborate below in a specific way in terms of a model Lagrangian.

It is generally understood that if kaons condensed at a density as low as $\sim 3n_0$,  the EoS of the matter would be considerably softened from the one without kaons condensed. This property was exploited by Brown and Bethe~\cite{brown-bethe} to argue that stars with mass $\gsim 1.5M_\odot$ would collapse and give rise to a considerably larger number of light-mass black holes in the Universe than estimated before~\cite{BLR}.  The recent observations of accurately measured $\sim 2$ solar-mass stars~\cite{1.97S} seemed to falsify this prediction and led to the conclusion that kaon condensation -- equivalently  hyperon presence  -- is ruled out.  We will argue that this conclusion is premature and unfounded. We note that there are at present a variety of proposals for loopholes to this conclusion, specifically in the presence of hyperons. It has been argued that under certain assumptions and with the adjustment of certain free parameters, the presence of hyperons can be accommodated  with $\gsim 2$ solar-mass stars.


In addressing the issue, two key questions arise. The first is how reliable the model, typically anchored on chiral Lagrangians, is for locating the critical density for kaon condensation, which is a phase transition.  In this connection, we note that all three ways of arriving at $n_K\sim 3n_0$ discussed in \cite{BLR} are based on the mean-field approximation of a given Lagrangian that encodes chiral symmetry.\footnote{The variety of approaches giving a similar critical density found in the literature wherein the boson sector and the fermion sector are treated separately are basically of the same approximation, although phrased in different ways.} What we will find is that if kaons condensed at as low a density as a few times the nuclear matter density, then the treatment for the condensation so far performed with the given Lagrangian could not be reliable. There can be at least one phase change, if not several, in the vicinity of the condensation phenomenon, most likely affecting the critical density importantly. One such phase change is of topological nature that is seen in the soliton-crystal description of dense matter -- on which a brief comment will be given below.

The second question is what happens, within the given Lagrangian framework, after kaons condense in baryonic matter. We are interested in exploring whether  under certain circumstances, the baryonic matter which is described as a Fermi liquid could not turn into a non-Fermi liquid with modified thermodynamic properties. Such a transition would consequently affect the EoS at a density relevant to the interior of compact stars.  We are motivated by a number of analogies in condensed matter systems that point to a need for  treatments that go beyond mean field theory, e.g., a renormalization group approach to the problem. For instance, a Fermi-liquid state coupled to scalar mesons, when the meson becomes critical -- that is, its mass tuned to zero -- turns to a non-Fermi-liquid state~\cite{kachru}. A similar transition is considered to occur when electrons in Fermi liquid are coupled to phonons from skyrmion crystal~\cite{watanabe}\footnote{In dense baryonic matter, a similar phenomenon is found to take place. It is found that the appearance of a half-skyrmion phase in skyrmion crystal at density $n_{1/2}\sim 2n_0$ drives a Fermi liquid to a state resembling non-Fermi liquid. More on this below.}. These phenomena involve anomalous dimensions. A particularly intriguing case that insinuates an analogy to what we are interested is the  space-time-dependent coupling between a fermion in Fermi liquid and a critical boson~\cite{silverstein}.\footnote{A possible implementation of this idea in our problem with the local density operator $\psi^\dagger (x)\psi (x)$ figuring in the coupling constant will be mentioned in the discussion section.}

Suppose that we arrive at a highly reliable theoretical tool -- which is yet to be found  -- for treating meson condensation in which we have strong confidence. If such a treatment still yielded a low kaon condensation density {\em and} made the EoS too soft to support $\gsim 2M_\odot$ stars within the standard theory of gravity,  then one could envisage a  possibility that a change in fundamental physics is required, with the EoS of compact stars providing a serious indication for a modified gravity. In fact, a suitably modified gravity such as for instance $f(R)$ gravity is argued to be able to  accommodate the massive stars even with the EoS softened by hyperons~\cite{odintsov}.

In this paper, we treat kaon condensation for kaons fluctuating around the Fermi-liquid fixed point arrived at in Wilsonian RG approach~\cite{shankar}. In compact-star matter, one is dealing with weak and chemical equilibrium, so the kaons figure via weak interactions. There the electron chemical potential increases as a function of density, and provides a crucial ``effective attraction'' to drive the kaon mass to drop.  For this process, the strong (attractive) KN-interaction and the electron chemical potential have to be treated self-consistently.

Here we focus on nuclear matter in strong interactions without weak and chemical equilibrium, as in \cite{LR2014}. Hyperons will not figure since they will be suppressed in the large $N$ limit -- where, as defined precisely below, $N\propto k_F$ where $k_F $ is the Fermi momentum. Once we understand what happens in nuclear matter, then we can address what could happen in compact-star matter. In compact stars, when the effective kaon mass drops sufficiently low so as to equal the electron chemical potential in compact-star matter, the electrons will beta decay to kaons and the kaons will then bose-condense~\cite{BTKR}. The kaon mass need not vanish for the condensation to set in. This means that the information that one gains from kaons in nuclear matter will give an upper bound for the critical density for condensation in compact stars.

 As a spin-off, there are other questions that call for answers. For instance, what happens to the baryonic matter in a Fermi liquid when the kaons condense? What form of state it is after condensation, specifically what is its EoS? What happens if a phase transition in the baryonic matter -- such as e.g., the skyrmion-half-skyrmion changeover to be mentioned below --takes place before or in the vicinity of the critical density at which kaons condense?
 These questions have never been addressed before. In this paper we make the first step to answering the questions.

It should be mentioned that the model we will analyze with the KN coupling defined by  Eqs.~(\ref{S1})-(\ref{couple1}) given below was first studied in  \cite{LRS}.  What differentiates this model from the usual chiral Lagrangian approach is that the treatment starts from the background defined by the Landau Fermi-liquid fixed point, instead of from the vacuum where the standard chiral perturbation approach is most likely applicable. This is along the line of the approach that exploits the notion of BR scaling~\cite{br91}. As is now understood, this notion is natural in the context of hidden local symmetry supplemented with broken scale symmetry. The premise on which our approach anchors is  a generalized hidden local symmetry Lagrangian for the meson sector endowed with a ``mended symmetry" that has the unbroken symmetry limit at high density in which the Goldstone $\pi$, scalar $s$,  and vectors $\rho$ (and $\omega$) and $a_1$ become massless. We shall call this ``mended HLS." The vector mesons $\rho$ (and $\omega$) and $a_1$ can be identified as emergent (hidden) local gauge fields and the scalar $s$ as the dilaton field of the spontaneously broken scale invariance at chiral restoration.

The approach we take is in the notion of the double decimation explained in \cite{BR:DD}. The Fermi-liquid matter to which kaons as pseudo-Goldstone bosons couple is the matter arrived at by the ``first RG decimation" from the chiral symmetry scale $\sim 4\pi f_\pi\sim 1$ GeV.

In \cite{LRS}, a KN interaction of the form Eq.~(\ref{couple1}), irrelevant in the sense of renormalization group (RG) due to the quadratic disperion relation for the kaon, was analyzed. It was seen there that the kaon mass, relevant classically, flows and how it  flows, as a signature for the instability  of the kaon-nuclear system, is governed by the parameter $h$ although it is irrelevant, and that  at one loop order, in certain parameter space, the flow necessarily leads to the vanishing of the ``effective kaon mass" signaling kaon condensation. The crucial issue there was the parameter space that delineates the critical line that  leads to or away from the condensation point. What happens to the Fermi liquid matter after the kaons are condensed was not analyzed in detail there but the immediate prediction was that the fermion chemical potential $\mu_F$ increased with a contribution proportional to $\la K\ra^2$ and since $\mu_F=k_F^2/2m^*$, for a given density (i.e., $k_F$),  the effective fermion mass $m^*$ decreased. The quadratic dispersion relation used there that resulted in approximating the effect of the electron chemical potential was, however, not correct for a Goldstone boson in medium -- which belongs to type-A~\cite{watanabe-murayama}, but the qualitative behavior was still valid.

In this paper we address the issue with a correct linear dispersion relation relevant for the kaon (as a pseudo-Goldstone boson) and a relevant KN interaction that can arise when the  HLS with mended symmetries is taken into account. We will find that the resulting theory, which is an expansion in $1/k_F$, differs basically from chiral perturbation theory where the expansion is in $k_F$. The prediction one can make is that kaons could condense at a density considerably lower than what has been predicted in ChPT.

\section{Setup of the Problem}
\subsection{Partition function}
We start by defining the action for our system.  In the standard approach anchored on chiral Lagrangians, one couples kaons as pseudo-Goldstone bosons to nucleons in baryonic matter treated either in standard nuclear many-body approach or in the mean-field approximation~\cite{BR-walecka,BLR} of, or in chiral perturbation theory with, the effective Lagrangians that are defined in the vacuum. Here we take a different starting point which is in line with the notion of sliding vacuum~\cite{BR:DD} in which nuclear matter arises as a fixed point, i.e., Fermi-liquid fixed point. This naturally incorporates the notion of BR scaling~\cite{br91} into the description of kaon condensation. Here we take the nuclear equilibrium state as the Fermi-liquid fixed point and couple (pseudo-) Goldstone mesons $\phi$ to quasiparticles $\psi$ in the vicinity of the Fermi-liquid fixed point. We write the partition function in Euclidean space as
\begin{equation}
Z = \int [d\phi][d\phi^\ast][d\Psi][d\Psi^\dagger] {\rm e}^{-S^E}
\end{equation}
\begin{eqnarray}
S^E&=& S_\psi^E + S_\phi^E +S_{\psi\phi}^E\ \label{S1}\\
S_\psi^E &=&\int  d\tau  d^3 \vec{x}
\, \Psi_\sigma^\dagger \left[ \partial_\tau + \epsilon \left( -i\vec{\nabla} \right) - \epsilon_F \right]\Psi_\sigma\nonumber\\
&& - \int d\tau  d^3 \vec{x} \, \lambda \Psi^\dagger_\sigma \Psi^\dagger_{\sigma^\prime} \Psi_\sigma\Psi_{\sigma^\prime} \label{fermion1}\\
S_\phi^E
&=&
\int d\tau  d^3 \vec{x} \phi^\ast \left( -\partial_\tau^2 - \vec{\nabla}^2 +m_\phi^2 +\cdots\right) \phi \label{boson1}\\
S_{\psi\phi}^E
&=&
- \int d\tau  d^3 \vec{x} \, h\, \phi^*\phi \Psi^\dagger_\sigma\Psi_\sigma, \label{couple1}
\end{eqnarray} where the sum over spin projection $\sigma$ (and $\sigma^\prime$) is implied.

We prefer to work in momentum space, and  so write
\begin{equation}
Z = \int [d\phi][d\phi^\ast][d\psi][d\bar{\psi}] { e}^{-\tilde{S}^E}
\end{equation}
with the actions
\begin{eqnarray}
\tilde{S}^E&=& \tilde{S}_\psi^E + \tilde{S}_\phi^E +\tilde{S}_{\psi\phi}^E\ \label{S2}\\
\tilde{S}_\psi^E &=&\int  \frac{d\epsilon  d^3 \vec{k}}{(2\pi)^4}
\,\bar{\psi}_\sigma \left\{ -i\epsilon +(e(\vec{k})-e_F)\right\} \psi_\sigma\nonumber\\
&& - \int \left( \frac{d\epsilon  d^3 \vec{k}}{(2\pi)^4}\right)^4 \lambda \bar{\psi}_\sigma \bar{\psi}_{\sigma^\prime} \psi_\sigma\psi_{\sigma^\prime} \delta^4(\epsilon,\,\vec{k})\label{fermion2}\\
\tilde{S}_\phi^E &=& \int \frac{d\omega d^3\vec{q}}{(2\pi)^4} \{\phi^*(\omega^2 +q^2)\phi +m_\phi^2\phi^*\phi+\cdots\} \label{boson2}\\
\tilde{S}_{\psi\phi}^E &=&-\int \left(\frac{d\epsilon d^3\vec{k}}{(2\pi)^4} \right)^2 \left(\frac{d\omega d^3 \vec{q}}{(2\pi)^4} \right)^2 \, h\,\phi^*\phi \bar{\psi}_\sigma\psi_\sigma \delta^4(\omega,\,\epsilon,\,\vec{q},\,\vec{k})\,. \label{couple2}
\end{eqnarray}
where $\psi$ and $\bar{\psi}$ are the eigenvalues of $\Psi$ and $\Psi^\dagger$ acting on $\left|\psi \right. \rangle$ and $\langle \left. \bar{\psi} \right|$,
\begin{equation}
\Psi \left|\psi \right. \rangle = \psi \left|\psi \right. \rangle \quad  {\rm and} \quad \langle \left. \bar{\psi} \right| \Psi^\dagger = \langle \left. \bar{\psi} \right| \bar{\psi}\,
\end{equation}
called fermion coherent state.
Although our notations for the fields are general as they can apply to other systems like pions/nucleons and electrons/phonons, we should keep in mind that we are specializing to the $K^-$ field for the boson and the proton and neutron for the fermion. We note that the action (\ref{S2}) is essentially the same model studied in \cite{LRS} except for the bosonic action. The bosonic action used there is different from (\ref{boson2}) in that the system considered there was assumed to be a compact-star matter in weak and chemical equilibrium (with the boson approximated to satisfy a quadratic dispersion relation -- which as mentioned, is most likely incorrect for kaons) whereas here we are dealing with relativistic bosons which will be tuned to criticality by attractive KN interactions.  It is interesting that a same-type action figures in condensed matter physics. In fact, (\ref{S2}) is quite similar to the action studied for quantum critical metals in \cite{kachru} from which we shall borrow various scaling properties for the renormalization group.

For later discussions, it is useful to identify the constant $h$ in (\ref{couple2}) with what is in the standard chiral Lagrangian. There are two terms contributing to it. In terms of the chiral counting, the leading term (${ \O} (p)$) is WT term and gives, for the s-wave kaon,
\be
h_{WT}\propto q_0/f_\pi^2
\ee
where $q_0$ is the fourth component of the kaon 4-momentum
and the next chiral order ($O (p^2)$) term is the KN $\Sigma$-term
\be
h_{\Sigma}\propto \Sigma_{KN}/f_\pi^2.
\ee

As written, the action for the fermion system (\ref{fermion2}) is for a Fermi liquid with marginal four-Fermi interactions~\cite{shankar}. So the energy-momentum of the quasiparticle (fermion) is measured with respect to the Fermi energy $e_F$ and Fermi momentum $k_F$. The bosonic action is for massive Klein-Gordon field, which later will be associated with a pseudo-Goldstone field, with higher-field terms ignored.  In contrast to that of the quasi-particle, the boson energy-momentum will be measured from zero.  We would like to look at meson-field fluctuations on the background of a fermionic matter given by the Fermi-liquid theory. The Fermi-liquid structure is expected to be valid as long as $N\equiv k_F/\bar{\Lambda} \gg 1$ where $\bar{\Lambda}=\Lambda-k_F$ where $\Lambda$ is the (momentum) cutoff scale from which mode decimation -- in the sense of Wilsonian -- is made.  For nucleon systems, the action (\ref{fermion2}) gives the nuclear matter stabilized at the fixed point, Landau Fermi-liquid fixed point, with corrections suppressed by $1/N$~\cite{shankar}. The four-Fermi interactions and the effective mass of the quasiparticle are the fixed-point parameters. We expect that as long as $N\gg 1$, the Fermi-liquid action can be trusted even at higher densities than $n_0$ {\it provided of course there are no phase transitions} that can destroy the Fermi-liquid structure.

For simplicity, we shall assume a spherically symmetric Fermi surface in which case we can set for the quasiparticle momentum
\be
\vec{k}=\vec{k}_F+\vec{l}\approx \hat{\Omega}(k_F+l).\label{fermionmomentum}
\ee
Then for $\bar{\Lambda} \ll k_F$, we have
\begin{equation}
e(k)-e_F \approx \vec{v}_F \cdot \vec{l} +O(l^2)
\end{equation}
where $v_F$ is the Fermi velocity $v_F=k_F/m^\star$ with $m^\star$ the effective quasiparticle mass.

\subsection{Scaling}\label{recipe}

\begin{figure}[ht]
\begin{center}
\includegraphics[width=7.0cm]{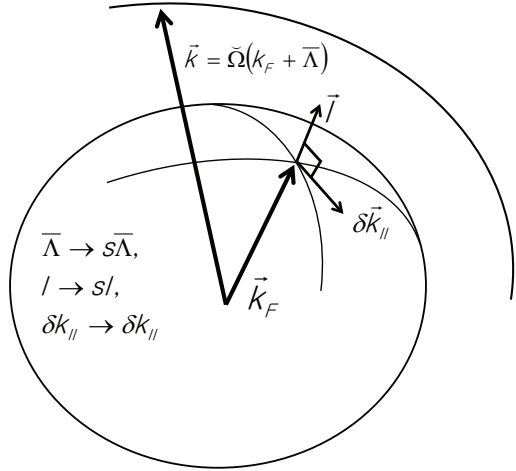}
\caption{Only the orthogonal component of the momentum to the fermi surface scales as $l \rightarrow s\, l$, where $\delta \vec{k}_{\parallel}$ is the parallel component of the momentum to the fermi surface.  }
\label{scaling}
\end{center}
\end{figure}
 Being an effective action, we need to set {the cutoff scale ${\Lambda}$} at which the classical action has bare parameters. Quantum effects are calculated by doing the mode elimination (``Wilsonian decimation") by lowering {the cutoff from ${\Lambda}$ to $s{\Lambda}$} with $s<1$. We shall do this closely following the procedure given by \cite{kachru}. The important point to note is that the boson and fermion fields have different kinematics. While the boson momentum is measured from the origin -- and hence the momentum cutoff is $\Lambda$ , the fermion momentum is measured from the Fermi momentum $k_F$. Therefore the mode elimination for the fermion involves lowering the fermion momentum from $\bar{\Lambda}=\Lambda-k_F$ to $s\bar{\Lambda}$. As noted above, the strategy in the fermion sector is to take the large $N$ limit where $N\equiv k_F/\bar{\Lambda}$.\footnote{ Note that the larger the $k_F$ or density, the more $\bar{\Lambda}$ shrinks, which would mean that the large $N$ argument would hold better in the fermion sector as the density increases. This suggests that the mean field approximation with effective Lagrangians -- chiral Lagrangian or hidden local symmetry Lagrangian -- would get better the higher the density. A caveat to this is that the argument would be invalid if there intervened non-perturbative phenomena such as the topology change suggested in the skyrmion crystal model of dense matter~\cite{LR-crystal}.}

 We define the scaling laws of fields and other quantities like the delta functions by requiring that the kinetic energy terms of the fermion and the boson be invariant under the scaling
\be
\epsilon\rightarrow s\epsilon, \ \omega\rightarrow s\omega, \ l \rightarrow sl, \ \vec{q}\rightarrow s\vec{q}. \label{vector}
\ee
As is seen from Fig.~(\ref{scaling}), only the fermion momentum orthogonal to {the Fermi surface} scales in the way the fermion energy and all components of the boson momentum do. Since the quasiparticle mass {$m^\star$}  is a fixed-point quantity, the Fermi velocity does not scale. Therefore we have the fields scaling as
\be
\phi&\rightarrow& s^{-3}\phi,\nonumber\\
\psi&\rightarrow& s^{-3/2}\psi\,, \label{field}
\ee for which we denote the scaling dimensions as $[\phi]=-3$ and $[\psi]=-3/2$\footnote{This notation will be used in what follows.} and the meson mass term scales
\begin{equation}
\int d\omega d^3q\, m_\phi^2 \phi^\ast \phi \rightarrow \int d\omega d^3q\, m_\phi^2 s^{-2} \phi^\ast \phi = \int d\omega d^3q\, m_\phi^{\prime\,2} \phi^\ast \phi,
\end{equation} with $m^\prime_\phi \equiv s^{-1} m_\phi$. This shows the well-known fact that the meson mass term is ``relevant."
Using the procedure of scaling toward the Fermi surface, we have
\be
[d\epsilon d^3k]=2,\  [\delta (\epsilon,k)] = -2\,. \label{fdelta}
\ee
This confirms that  the four-Fermi interaction term in (\ref{fermion2}) is marginal.\footnote{For two or three spatial dimensions, this applies to only forward scattering and BCS-type scattering. Others are marginal.}

The scaling of the coupling term (\ref{couple2}) is more subtle. Using the scaling dimensions
\be
[\phi]=-3, \ [\psi]=-3/2, \ [d\epsilon d^3\vec{k}]=2, \ [d\omega d^3\vec{q}]=4,\label{scaling-couple}
\ee
we find the scaling of the integrand $I_{\psi\phi}$ of the action (\ref{couple2})  written as $\int I_{\psi\phi}$ is
\be
[I_{\psi\phi}]=3+[h]+[\delta (\omega,\epsilon,\vec{q},\vec{k})],\label{Iscaling}
\ee
where the bare coupling constant $h$ will have the scaling dimension $[h_{WT}]=[q_0]=1$ and $[h_{\Sigma}]=0$. The $\psi\phi$ coupling will be ``relevant" if  $[I_{\psi\phi}]<0$, marginal if $=0$ and ``irrelevant" if $>0$. It is thus the scaling of the delta function that determines the scaling of the coupling $h$. In Appendix A, we suggest that depending upon the kinematics of the nucleon momentum parallel to {the Fermi surface}, there can be two scaling possibilities
\be
[\delta (\omega,\epsilon,\vec{q},\vec{k})]=-2, \label{case1}
\ee
and
\be
[\delta (\omega,\epsilon,\vec{q},\vec{k})]=-4. \label{case2}
\ee
As in Appendix A, we refer to (\ref{case1})  as ``case 1" and (\ref{case2}) as ``case 2".
From (\ref{Iscaling}), we get for the two cases of possible scaling:
\begin{enumerate}
\item {\bf Case 1}

\be
[I_{\psi\phi}]_{h_{WT}}=2, \  [I_{\psi\phi}]_{h_{\Sigma}}=1.\label{I}
\ee
Both couplings are irrelevant.
\item {\bf Case 2}

\be
[I_{\psi\phi}]_{h_{WT}}=0, \  [I_{\psi\phi}]_{h_{\Sigma}}=-1.\label{II}
\ee
Here the Weinberg-Tomozawa coupling is marginal, while the $\Sigma$-term coupling is relevant.
\end{enumerate}
A priori, there is no obvious reason to pick one case over the other. One way of seeing how these two different cases arise is to integrate over the (initial or final) fermion or boson using the delta function in the action (\ref{couple2}).  How one does the integration gives two different results. If one integrates over the fermion, from (\ref{scaling-couple}),  one gets the case 1 whereas integrating over the boson gives the case 2. This difference must therefore be tied to the difference in the way the momentum decimation is done, i.e., the scaling toward the Fermi surface (fermion) or to the origin (boson).

There are two (perhaps related) mechanisms that lead to the case 2 (\ref{II}).  We will argue in Section \ref{hls} that within the precisely defined framework in which the problem is formulated, the case 2 -- (\ref{case2}) with (\ref{II}) -- is favored. This is discussed in terms of a mended HLS Lagrangian. Another reasoning -- which is possibly connected to the first mechanism via BR scaling -- is mentioned in the discussion section.

\subsection{Hidden local symmetry with scale invariance}\label{hls}
Here we present an argument that hidden local symmetry supplemented with scale invariance could lead to the scaling (\ref{II}).

When the non-linear sigma model, with chiral perturbation theory as the flagship effective field theory for nuclear dynamics, is elevated to hidden local symmetry  theory with the vector mesons (i.e., $\rho$ and $\omega$ for two-flavor case) figuring as local gauge fields~\cite{HY:PR},  it acquires the power, absent in the non-linear sigma model, to make certain unique predictions for the properties of hadrons in the environment in which the vacuum is strongly modified by temperature or density. One striking feature -- which however remains neither confirmed nor falsified by experiments -- is that when the quark condensate vanishes in the chiral limit as expected at high temperature or density, what is referred to as ``vector manifestation (VM)" fixed point is approached with the vector meson masses vanishing. If one assumes $U(2)$ flavor symmetry for the vector mesons, this then means both $\rho$ and $\omega$ would be massless at the VM.

In addition, if a scalar degree of freedom corresponding to the (pseudo-)Goldstone boson of broken scale symmetry, say, ``dilaton", is implemented, then the argument based on ``mended symmetries"~\cite{weinberg} would imply that the dilaton mass will also approach zero at the chiral restoration point. Let us call this ``mended hidden local symmetry (mHLS)" and the approach to the vanishing masses ``mended vector manifestation (mVM)."

In mHLS, one can think of the WT-type term and the $\Sigma$-type term in (\ref{couple2}) arising in the vacuum  when the vector mesons and the scalar are, respectively, integrated out, thereby giving local interactions.\footnote{These terms in HLS implemented with dilaton, i.e., mHLS, have the same forms as the WT and $\Sigma$ terms of the chiral Lagrangian, so we loosely call them by the same names.}  This localization is clearly justified for low-energy dynamics in the vacuum given the heavy masses involved.  In dense medium, however, this localization would become problematic if the density (or temperature) were in the vicinity of the chiral restoration point where the VM phenomenon sets in.

What can we say about the RG flow of dense nuclear matter in mHLS?

It was found in \cite{maetal,WGP} that the flavor $U(2)$ symmetry for the vector mesons ($\rho, \omega$) -- which is a fairly good symmetry in the vacuum -- is strongly broken in medium. As a consequence,  while the mass of $\rho$ meson is to decrease in accordance with the approach to the VM fixed point, the $\omega$ mass must remain more or less unaffected by the density except, perhaps, near the chiral restoration point. Though not firm, this feature is not inconsistent with  what is observed with the EoS of compact-star matter~\cite{dongetal}. This means that the $\omega$ meson-exchange contribution to the Weinberg-Tomozawa term in medium will be the same as the local form in the vacuum. As for the $\rho$ contribution, the dropping $\rho$ mass could make the local form a suspect at high density but its contribution is proportional to $\frac 13(n_p-n_n)$ -- compared to $(n_p+n_n)$ for the $\omega$ exchange -- where $n_{n,p}$ is (proton, neutron) number density. For $n_n>n_p$ as in compact stars, the contribution is repulsive. Therefore hidden local fields are expected not to modify the RG properties of the coupling (\ref{couple2}) in medium.

 The situation with the $\Sigma$-like term contribution, however, can be quite different if the scalar dilaton implemented into HLS Lagrangian drops in mass as is indicated in \cite{maetal}. It is not clear whether the behavior of the scalar dilaton can be incorporated  into a mVM structure. But ``mended symmetries''~\cite{weinberg} for the $\pi$, $\rho$, scalar (denoted $\epsilon$ in \cite{weinberg})  and $a_1$ do suggest that the scalar mass could also drop significantly at high density. We shall refer to the approach to mended symmetries as ``mended-symmetry limit."

Implementing satisfactorily the mVM properties mentioned above would require a consistent treatment of both hidden local symmetry and conformal symmetry which are intricately linked to each other. We have at present no systematic treatment of the two symmetries in dense medium but it is possible to gain insight into what could be going on by tree-level consideration of the Feynman diagrams in mHLS.

The argument is quite simple. The crucial point to note is that in the one-$\phi$ exchange tree diagrams, the KN coupling giving rise to (\ref{couple2}) differs from that leading to the four-Fermi interaction (\ref{fermion2})  by that one of the two $\psi\psi\phi$ vertices in the former is replaced by a $KK\phi$ vertex. Within the scaling rule we adopt, the $\psi\psi\phi$ vertex is marginal but the $KK\phi$ vertex is relevant. All other quantities are the same. Since in our approach, we want the four-Fermi interaction to remain marginal so that the baryonic matter remains in Fermi liquid, we need to enforce that the presence of the $\phi$ field leave unaffected the marginal four-Fermi interaction. As shown in Appendix \ref{one-meson}, this is achieved by assigning a definite scaling property to the $\phi$ propagator. By imposing this scaling condition on the $\phi$ propagator that figures in the KN-KN interaction, one sees immediately that the $\Sigma$-term type KN coupling in (\ref{couple2}) must be relevant. This is simply because the $KK\phi$ vertex is relevant. One can verify this result by an explicit counting of the scalings involved in the Feynman diagrams as shown in Appendix \ref{one-meson}.

Now using the same simple counting rule, it is easy to see that the WT term should be marginal, i.e., it is in the case 2 (\ref{II}). For this, simply replace the $\psi\psi\phi$ vertex of the scalar-exchange term by a $\psi\psi\omega$ coupling and the $KK\phi$ vertex by a $q_0 KK \omega$ vertex, which amounts to replacing a marginal coupling by a marginal coupling and a relevant coupling by a marginal coupling. The net result is marginal because $q_0 KK\omega$ vertex is marginal.
\section{Renomalization Group Equations}
We are interested in the flows of the kaon mass and the kaon-nucleon coupling and the effect of kaon condensation on the Fermi liquid structure of the baryonic system.
Since the WT coupling in its local form is irrelevant at the tree level for the case 1 and at one-loop level for the case 2, we focus on the $\Sigma$-term coupling. Depending on the two cases of the delta function scaling, its tree order scaling is either irrelevant or relevant. What matters crucially is then the loop correction, so we first look at one-loop contributions with the $\Sigma$-term coupling. Two-loop corrections should be suppressed for the large $N$ limit.

One-loop corrections to the beta function for the WT coupling are given in the next subsection.
\subsection{$\Sigma$-like coupling}
\subsubsection{Tree-order scaling}
As shown above, under the scale transformation the coupling, $h$ in the action scales
\begin{equation}
h^\prime = s^{-a_h}\,h\,,
\end{equation}
where $a_h=-1 (+1)$ for the case 1 (case 2). The kaon mass scaling is relevant
\be
m_\phi^{\prime\, 2}=s^{-2} m_\phi^2.
\ee
\subsubsection{One-loop-order scaling}
\begin{figure}[ht]
\begin{center}
\includegraphics[width=8.0cm]{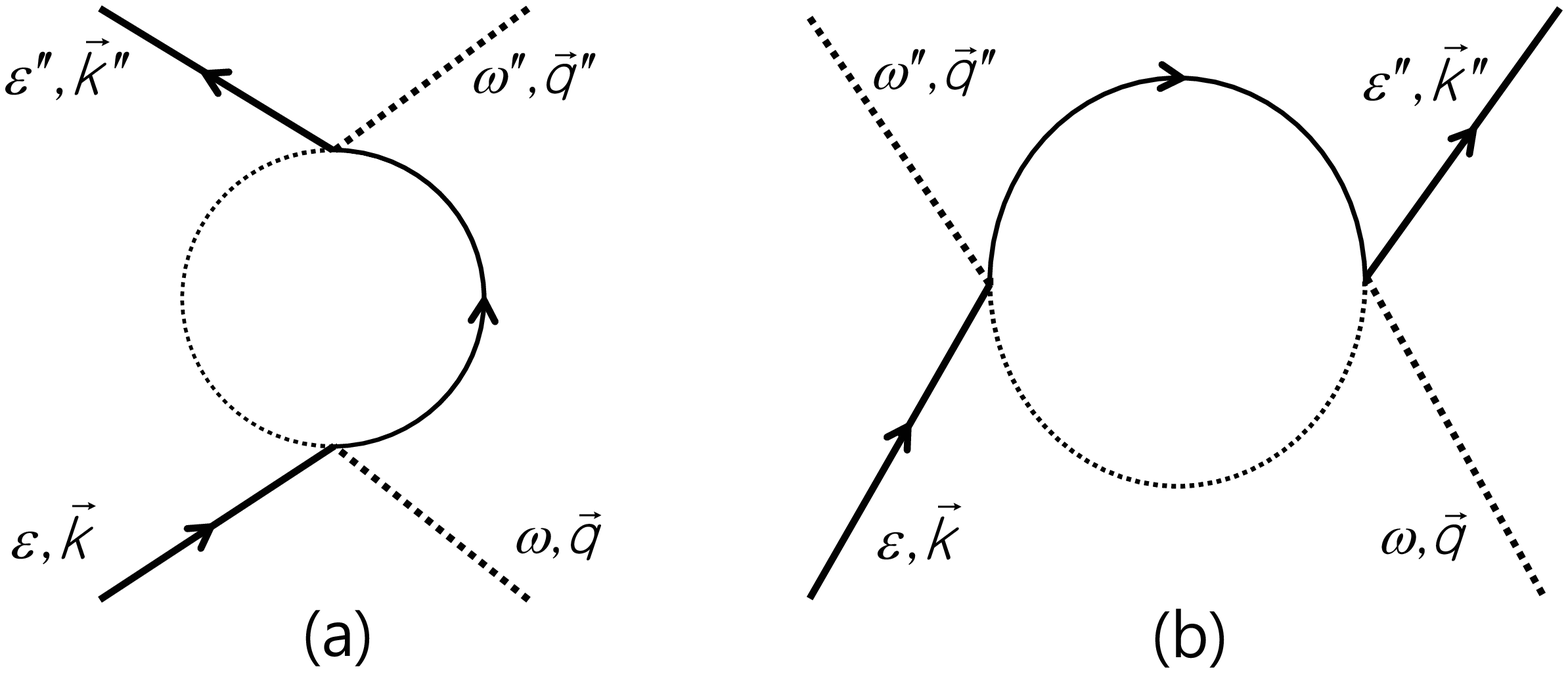}
\caption{One-loop graphs contributing to $h$. The process corresponds to f(ermion) ($\epsilon,\vec{k}$)+ m(eson) ($\omega,\vec{q}$)$\rightarrow$  f(ermion) ($\epsilon^{\prime\prime},\vec{k}^{\prime\prime}$)+ m(eson) ($\omega^{\prime\prime},\vec{q}^{\prime\prime}$).}
\label{kaon_fermion}
\end{center}
\end{figure}
\begin{figure}[bh]
\begin{center}
\includegraphics[width=5.0cm]{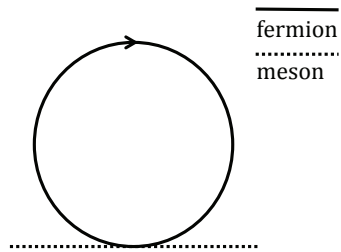}
\caption{One-loop graph contributing to $m_\phi$.}
\label{one_loop_mass}
\end{center}
\end{figure}
The one-loop graphs contributing to $h$ and $m_\phi$ are given, respectively, in Figs. \ref{kaon_fermion} and \ref{one_loop_mass}. {\it What is significant for our analysis is that the one-loop graphs Figs.~\ref{kaon_fermion} turn out not to contribute to $h_\Sigma$.} Since this result is nontrivial, we sketch here the calculation for the vanishing contribution to $h_\Sigma$ with  the details relegated to an appendix. The calculation for $h_{WT}$ -- which does not vanish -- is given in the next subsection.

 For the kinematics involved, the figures (a) and (b) of Fig.\ref{kaon_fermion} give the same result, so we focus on figure (a).
Denote the loop energy-momentum of the fermion and of the meson, respectively, as  $(e^\prime, \vec{k}^\prime)$ and $(\omega^\prime, \vec{q}^\prime)$.  Then
the energy-momentum conservation at each vertex leads to
\begin{eqnarray}
  \omega^\prime &=&  \epsilon^\prime -Q_0\,, \\
 \vec{q}^{\,\prime} &=&  \Omega\left(K_F + l^\prime \right) - \Omega\left(K_F + l \right)\,,
\end{eqnarray}
where $Q_0 \equiv \epsilon - \omega$ and $\Omega\left(K_F + l \right) \equiv \vec{k} - \vec{q}$.  Note that $Q_0$ is the external energy difference between the fermion and the meson satisfying the dispersion relations,{
\begin{eqnarray}
\epsilon &=& \frac{\vec{k}^{\, 2}}{2m^\star} -\epsilon_F\,, \\
\omega^2 &=& \vec{q}^{\, 2} + m_\phi^2\,.
\end{eqnarray}}
Then the one-loop contribution to the action (\ref{couple2}) from Fig.~\ref{kaon_fermion}(a) -- detailed in Appendix B -- is given by
\begin{eqnarray}
\delta S^{(a)}
&\equiv&
{} \int \frac{\left(d\omega d^3q \right)^2 \left( d\epsilon d^3k \right)^2}{(2\pi)^{16}} h^2 \phi^\ast_< \phi_< \bar{\psi}_< \psi_< \, \delta^4 \left( \epsilon, \omega, \vec{k}, \vec{q}\right)
\nonumber\\
&&
{} \times \int \frac{d\epsilon^{\prime} d^3 k^{\prime}}{(2\pi)^4} \frac{-1}{i\epsilon^{\prime} - \vec{v}_F \cdot \vec{l}^\prime} \, \frac{1}{ \left(\epsilon^\prime - Q_0\right)^2 + \left( l^{\prime} - l \right)^2 + 2k_F^2\left( 1 - \cos \theta \right) + m_\phi^2 }\,
\end{eqnarray}
where fields with $<$ denote low-frequency modes retained after integrating out high-frequency modes as defined in Appendix B.
Since the Renormalization Group Equation(RGE) should not depend on the external kinematics, i.e.,  $Q_0$ and $l$, one can set $Q_0=l=0$ to simplify the integral. One can easily convince oneself that the integral vanishes, as shown explicitly in Appendix \ref{B}. Thus there is no one-loop contribution to the $\beta$ function for $h_\Sigma$.

The one-loop graph Fig.~\ref{one_loop_mass} contributing to the $\phi$ mass is easier to evaluate. By decimating from $\bar{\Lambda}$ to $s\bar{\Lambda}$ in the nucleon loop, we get
{\begin{eqnarray}
\int \frac{d \omega d^3\vec{q}}{(2\pi)^4} \phi^\ast \phi s^{-2}\left[ -m_\phi^2 +\frac{\gamma h k_F^2 }{2\pi^2} \int_{s\bar{\Lambda} < \left|l^\prime \right| < \bar{\Lambda}} dl^\prime\, \left|{\rm sgn}\left( l^\prime \right) \right|\right]\,,
\end{eqnarray}
where $\gamma$ is the degeneracy factor for flavor (=2) and spin (=2): $\gamma=4$ for nuclear matter and $\gamma=2$ for neutron matter. It follows from above that
\begin{equation}
\delta \left[m_\phi^2 \right] = s^{-2} \left[ m_\phi^2 - \frac{\gamma h \bar{\Lambda} k_F^2 }{\pi^2} \left( 1-s \right) \right] - m_\phi^2\,.
\end{equation}}

\subsubsection{Flow analysis}
Although the case 2 (\ref{II}) is favored by out approach based on mHLS, we treat both cases for comparison.

From what's discussed above, it is straightforward to write down the RGEs. Setting $t\equiv -\ln s$,
\begin{eqnarray}
\frac{d\, m_\phi^2}{dt} =  2\, m_\phi^2 - A \, h\,, \label{massrge}\\
\frac{d\, h }{dt} =  a_h h - B\, h^2\,, \label{hrge}
\end{eqnarray} where{
\begin{eqnarray}
A &=& \frac{\gamma k_F^2 \bar{\Lambda}}{\pi^2}\,, \\
B &=& 0 \,.
\end{eqnarray}}
It is easy to get the analytic solutions for $m_\phi$ and $h$
\begin{eqnarray}
m_\phi^2 (t) &=& \left( m^2_\phi(0) - \frac{h(0) A}{2-a_h} \right) {\rm e}^{2t} + \frac{h(0) A}{2-a_h} {\rm e}^{a_h t}\,,\label{mphi}\\
h(t) &=& h(0) {\rm e}^{a_h t}\,,\label{hterm}
\end{eqnarray} where $a_h =\pm 1$ for the case 2 (relevant)/case 1 (irrelevant).

Given the analytic solution, we can work out how $m_\phi^2$ and $h$ flow as $t$ increases ($s$ decreases) for given values of $A$, $h(0)$ and $m_\phi^2(0)$.
From Eqs.~ (\ref{mphi}) and (\ref{hterm}), we have the relation,
\begin{equation}
\frac{m_\phi^2(t) - \frac{A\,h(t)}{2-a_h}}{\left[ h(t) \right]^{2/a_h}} = \frac{m_\phi^2(0) - \frac{A\,h(0)}{2-a_h}}{\left[ h(0) \right]^{2/a_h}}\,, \label{rel_c1}
\end{equation} which is satisfied for any value of $t$. It is convenient to introduce the new parameter $c_{a_h}$ which is independent of $t$,
\begin{equation}
c_{a_h} \equiv \frac{m_\phi^2(0) - \frac{A\,h(0)}{2-a_h}}{\left[ h(0) \right]^{2/a_h}}\,. \label{rel_c2}
\end{equation}
The quantities at $t=0$,  i.e., $m_\phi^2(0)$ and $h(0)$, are given at $s=1$, that is, at the scale from which the decimation starts for a given $k_F$. {\it These parameters in the bare action -- and the parameter $c_{a_h}$ -- depend only on density. Since the flow depends on $c_{a_h}$, the RG properties of dense medium depend on the density dependence of the parameters of the bare chiral Lagrangian at the scale  ${\Lambda}$. This is equivalent  to the BR scaling~\cite{br91} that is obtained in relativistic mean field treatment of effective Lagrangians of the HLS-type which is again equivalent to  Landau Fermi-liquid theory.}

\begin{figure}[ht]
\begin{center}
\includegraphics[width=16.0cm]{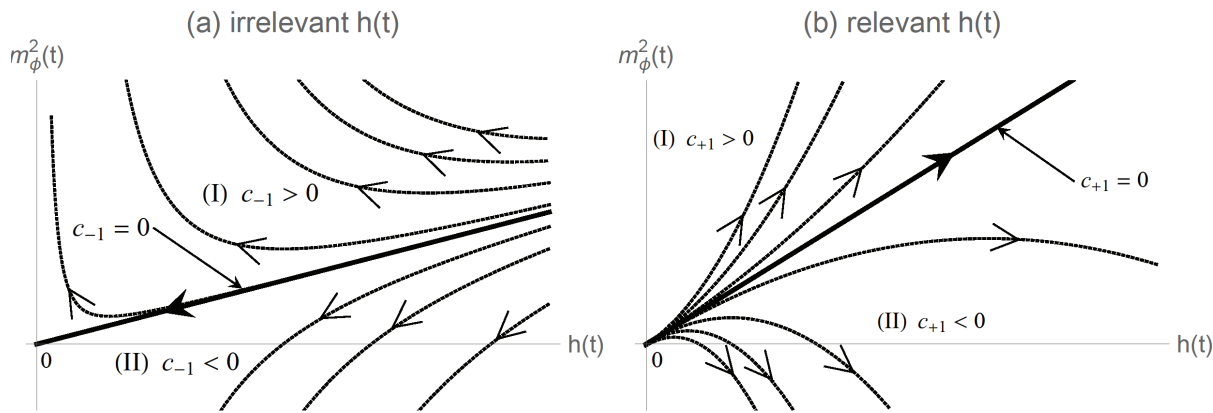}
\caption{RG flows of the parameters, $m_\phi^2(t)$ and $h(t)$, are shown in the parameter space. The lines are drawn with the fixed value of A and the different values of $c_{a_h}$. The black solid line stands for the ``critical line (or surface)" that delineates the parameter spaces. }
\label{RG_flow}
\end{center}
\end{figure}
Now using Eqs.~(\ref{rel_c1}) and (\ref{rel_c2}), we readily obtain the formulas for $m_\phi (t)$ for the two cases $a_h=\pm 1$
\be
m_\phi^2(t) &=&  \frac{c_{-1}}{h(t)^2} + \frac{A}{3}h(t) \ \ \ {\rm for}\ \ a_h=-1,\\
&=&  c_{+1}h(t)^2 + A\,h(t) \ {\rm for}\ \ a_h=+1.
\ee
The flows $m_\phi (t)$ vs. $h(t)$  are plotted in Fig.~\ref{RG_flow} for both $a_h=-1$ (left panel) and $a_h=1$ (right panel) for various values of $c_{a_h}$ that depend on $k_F/\bar{\Lambda}$. We note that $m_\phi^2(t)$ flows toward zero and becomes negative for $c_{- 1} \leq 0$ and $c_{+ 1} < 0$. This signals the instability toward kaon condensation. What happens after the condensation is a matter that goes beyond the model. It could be stabilized by terms higher order in the kaon field not taken into account in our analysis.

The conclusion is that only the $m_\phi(0)$ and $h(0)$ that  satisfy
\be
m_\phi^2(0) &\leq& \frac{A}{3} h(0), \ \ {\rm for} \ \ \ a_h=-1\, \label{cond1kc}\\
m_\phi^2(0) &<& A h(0) \ \ {\rm for} \ \ \ a_h=+1 \label{cond2kc}
\ee
will be in the parameter space for the condensation to take place. The value $c_{\pm1} = 0$ defines the critical line in the $h(t)-m_\phi^2(t)$ plane, any point on which the parameters flow toward (for $a_h=-1$) or from (for $a_h=+1$) the origin satisfying
\be
m_\phi^2(t) &=& \frac{A}{3} h(t) \ \ {\rm for} \ \ \ a_h=-1,\\
m_\phi^2(t) &=& A\, h(t) \ \ {\rm for} \ \ \ a_h=+1.
\ee
Note that in the RG flows of $m_\phi(t)$ and $h(t)$,  as $t$ goes up, the attractive KN interaction gets weaker if $a_h = -1$  and gets stronger if $a_h = +1$.

\subsubsection{Critical densities}
We are unable to precisely pin down the critical density for kaon condensation from the one-loop RG analysis given above. We can, however, use the critical line to get the ``parameter window''  that signals the instability toward the condensation.

At some density, according to the values of $c_{a_h}$, the RG flows of $m_\phi^2 $ and $h(t)$ will follow certain of the flow lines in Fig.~\ref{RG_flow}.
If the value of $c_{a_h}$ is changed by changing $k_F$ to $k_F^\prime$, the RG flow of $m_\phi^2$ and $h(t)$ follows from the line (in $k_F$) to the line (in $k_F^\prime$).
By locating the sign change of
\begin{equation}
m_\phi^2(0)- \frac{A\,h(0)}{2-a_h}\,,\label{kaonmass}
\end{equation}
one can then locate the density at which the phase is moved to the region that signals the instability toward the kaon condensation. We use Eqs.~(\ref{cond1kc}) and (\ref{cond2kc}) to make this estimate. Note that this does not pinpoint the critical condensation density. The sign change of $c_{a_h}$ takes place when
\begin{equation}
n\equiv \gamma\frac{k_F^3}{6\pi^2} = \frac{N(2-a_h)}{6}\frac{m_\phi^2(0)}{h(0)}.
\end{equation}
If one takes $h(0)=\Sigma_{KN}/f^2$ from chiral Lagrangian, we get
\be
n\approx \frac{(2-a_h)N}{6}\frac{m_K^2 f^2}{\Sigma_{KN}} \label{nkaon}
\ee
where $f\approx f_\pi$. For ${\cal H}\equiv \frac{(2-a_h)N}{6}\approx 1$, this result reproduces the tree-order critical density in chiral perturbation theory~\cite{LBMR}. Just for illustration, let us take $m_\phi(0)\approx 500$ MeV and $h(0) \approx  \Sigma_{KN}/f_\pi^2\approx 250\, {\rm MeV}/(93\, {\rm MeV})^2$\footnote{The value for $\Sigma_{KN}\approx 250$ MeV is the upper bound obtained in lattice QCD calculations~\cite{sigmaterm}.}. Then, by increasing $k_F$, we move from the flow lines with $c_{a_h} > 0$  toward the flow line with $c_{a_h} < 0$. In this way, we obtain the parameter window for the condensation density $n_K\sim 6n_0$ for $\frac{(2-a_h)N}{6}\approx 1$. In the density above $n_K$, we satisfy the conditions Eqs.~(\ref{cond1kc}) and (\ref{cond2kc}), and $m_\phi^2(t)$ will flow to zero as $t$ increases.

As mentioned, the RG analysis anchored on the Fermi-liquid fixed point relies on $N=k_F/\bar{\Lambda}$ being large. Thus the higher the density, the better the one-loop approximation becomes, with higher-loop terms suppressed for $N\gg 1$. The quantity ${\cal H}\equiv \frac{(2-a_h)N}{6}\approx 1$ implies $N\sim 6$ for the case where the KN coupling is relevant, i.e.,  $a_h=+1$, whereas for the irrelevant case with $a_h=-1$, $N\sim 2$.  This implies that a relevant coupling (which we suggested is possible in HLS theory supplemented with a dilaton) would make the large-N arguments work more powerfully for arriving at a low condensation density.

The upshot of the above remark is that the relevance of the $\Sigma$-term scaling brings a large reduction of the critical density, say, by a factor of 3, from the naive irrelevant coupling expected in the absence of mHLS. Equally importantly, if we consider kaon condensation in compact-star matter which is in weak and chemical equilibrium, incorporating $\mu_e$ will reduce further the critical density from  what is estimated in nuclear matter. In compact-star matter, chemical equilibrium relates $\mu_K$ to $\mu_e$ which can be as large as $\sim m_K/2$.  The consideration of a large kaon chemical potential might even modify the dispersion relation of the kaon field as noted in \cite{LRS}. This is a future work to be done.

\subsection{Weinberg-Tomozawa-type coupling}
The Weinberg-Tomozawa-type coupling has the meson fourth momentum $q_0$ appearing in $h$. This makes the one-loop graph contribution to $h_{WT}$ non-vanishing contrary to what happened for the $\Sigma$-term coupling.
For simplicity, we take the action for the Weinberg-Tomozawa-type interaction in the form
\begin{eqnarray}
\tilde{S}_{WT}^E =-\int \left(\frac{d\epsilon d^3\vec{k}}{(2\pi)^4}\right)^2 \left(\frac{d\omega d^3 \vec{q}}{(2\pi)^4}\right)^2 \, ih_{WT}\left(\omega + \omega^{\prime\prime} \right)\phi^*(\omega,\, \vec{q})\phi(\omega^{\prime\prime},\, \vec{q}^{\,\prime\prime}) \bar{\psi}_\sigma\psi_\sigma \delta^4(\omega,\,\epsilon,\,\vec{q},\,\vec{k})\,. \label{coupleWT}
\end{eqnarray}
In doing the loop calculation similarly to the diagrams in Fig.~\ref{kaon_fermion},
we get
\begin{eqnarray}
\delta S
&\equiv&
{} \int \frac{\left(d\omega d^3q \right)^2 \left( d\epsilon d^3k \right)^2}{(2\pi)^{16}} h_{WT}^2(\omega + \omega^\prime) \phi^\ast_< \phi_< \bar{\psi}_< \psi_< \, \delta^4 \left( \epsilon, \omega, \vec{k}, \vec{q}\right)
\nonumber\\
&&
{} \times \int \frac{d\epsilon^{\prime} d^3 k^{\prime}}{(2\pi)^4} \frac{-1}{i\epsilon^{\prime} - \vec{v}_F \cdot \vec{l}^\prime} \, \frac{\epsilon^\prime}{ \epsilon^{\prime\,2} + l^{\prime\,2} + 2k_F^2\left( 1 - \cos \theta \right) + m_\phi^2 }\,,
\\
&\approx&
{} \int \frac{\left(d\omega d^3q \right)^2 \left( d\epsilon d^3k \right)^2}{(2\pi)^{16}} h_{WT}^2(\omega + \omega^\prime) \phi^\ast_< \phi_< \bar{\psi}_< \psi_< \, \delta^4 \left( \epsilon, \omega, \vec{k}, \vec{q}\right)
\nonumber\\
&&
{} \times \int \frac{d\epsilon^{\prime}\, k_F^2\, d l^{\prime} d^2 \Omega}{(2\pi)^4} \frac{-1}{i\epsilon^{\prime} - \vec{v}_F \cdot \vec{l}^\prime} \, \frac{\epsilon^\prime}{ \epsilon^{\prime\,2} +l^{\prime\,2} + m_\phi^{\prime\, 2} }\,,
\end{eqnarray} which gives
\begin{equation}
-\int \left(\frac{d\epsilon d^3\vec{k}}{(2\pi)^4}\right)^2 \left(\frac{d\omega d^3 \vec{q}}{(2\pi)^4}\right)^2 \, ih_{WT}\left(\omega + \omega^{\prime\prime} \right)\phi^*(\omega,\, \vec{q})\phi(\omega^{\prime\prime},\, \vec{q}^{\,\prime\prime})\bar{\psi}\psi \delta^4(\omega,\,\epsilon,\,\vec{q},\,\vec{k}) + 4\delta S.
\end{equation} After doing the integration, we have for the case 1 (case 2)
\begin{equation}
\frac{d\,h_{WT}}{dt} = -2({\rm or}\, 0) h_{WT} - \frac{4k_F^3 B^\prime}{N(2\pi)^3}h_{WT}^2, \label{RGE_WT}
\end{equation} where
\begin{equation}
B^\prime = \int d^2\Omega \frac{1}{v_F \bar{\Lambda} + \sqrt{\bar{\Lambda}^2 + m_\phi^2 + 2k_F^2\left( 1 - \cos \theta \right)}} \,.
\end{equation}
Although $h_{WT}$ has the non-vanishing one-loop contribution,
suppressed by $1/N$ , it makes $h_{WT}$ more irrelevant for the case 1. As for the case 2 that we argue to be the correct scenario,  since the tree-order coupling is marginal and the one-loop correction is irrelevant, it is {\it marginally irrelevant}. Note also that $B^\prime$ is non-negative, so there is no non-trivial fixed point in $h_{WT}$.  We remark below what the implication of this marginally irrelevant term is in kaon condensation is.
\section{Discussions and Concluding Remarks}
We briefly recapitulate what we have found in this work. When the kaon condensation is treated  on the {\it background of Fermi liquid varying with density} in the framework of chiral Lagrangians, the Weinberg-Tomozawa-like term that plays a predominant role for kaon-nuclear interactions at low energy associated with the $\Lambda (1405)$ resonance is found to be at best marginally irrelevant in the RG sense at high density and hence unimportant for bringing the system to kaon condensation. It is the $\Sigma_{KN}$-like term which is subleading in chiral perturbation theory, interpreted in terms of the dilaton scalar figuring in  BR scaling,  that can trigger kaon condensation at densities relevant to compact stars, i.e. $n_K \lsim 4n_0$. That the $\Lambda (1405)$ resonance -- which is driven by the WT term  in chiral effective field theory -- plays no important role in the condensation {\it per se} was already foreseen many years ago even in standard chiral perturbation calculations starting with a chiral Lagrangian defined in the vacuum~\cite{LBMR}. The RG analysis of this paper confirms what was found there in terms of the derivative coupling that is controlled by chiral symmetry.  As in low-order chiral perturbation theory, the critical density goes like $\sim 1/\Sigma_{KN}$ in our RG analysis of the parameter window. Though the analysis does not allow us to pin down the critical condensation density $n_K$,  it clearly indicates that the Fermi-liquid fixed point structure as arising in the mended-symmetry-implemented model gives the “relevant” RG coupling for the $\Sigma_{KN}$-type interaction. This is how the RG approach anchored on the Fermi liquid structure would predict a considerably lower, say, a factor of $\simeq$ 3, critical density than that of ChPT. Furthermore as the scalar mass approaches the mended-symmetry limit, the critical density proportional to $m_s^2$(scalar mass squared) given in mHLS -- or $f^2$ in the eq. (\ref{nkaon}) -- will be lower than the density without mended symmetries.

The apparently insignificant role that the leading-chiral-order (Weinberg-Tomozawa) term for KN interactions plays for the condensation process {\it per se} raises a tension with what is found in anti-kaon nuclear interactions extensively studied both experimentally and theoretically. It has been shown quantitatively~\cite{weise} that the coupled-channel approach with the driving term given by the Weinberg-Tomozawa term captures well kaon-nuclear interactions at low energy. In this approach, the $\Lambda (1405)$ resonance controlled by the WT term plays a crucial role. In fact, this is the approach adopted in most of the theoretical works on low-energy kaon-nuclear physics found in the literature. Furthermore when $K^-$'s are present, bound, in dense medium,  the symmetry energy, an important ingredient for the EoS of compact stars, of the system in chiral perturbation theory at tree order is found to have a significant additional term proportional to the square of the bound-kaon wave function determined more or less entirely by the WT term ~\cite{LR2014}.
In both cases mentioned above, there is no visible direct role for the $\Sigma$-term coupling. What transpires here then is that the critical density is chiefly governed by the chiral symmetry breaking term while low-energy kaon-nuclear dynamics and the kaonic effect on the EoS of dense baryonic matter are controlled by the chirally symmetric term. This could be better understood if the WT term, although playing no (significant) role in determining the critical density $n_K$, does figure in the EoS at the condensation point. A clear understanding of this state of matter could be gained when the generalized HLS model is treated self-consistently in weak and chemical equilibrium appropriate for compact stars. We return to this matter in a future publication.

It has been recently suggested, based on a skyrmion-crystal description of dense matter~\cite{LR-crystal}, that  there can be  a topological phase change at a density $n\gsim 2n_0$ which would make the popular mean-field approximation used to describe the EoS of dense matter breakdown~\cite{WGP}. Since the mean-field treatment of nuclear effective Lagrangians is equivalent to Landau-Migdal Fermi liquid theory~\cite{matsui}, this would imply that the Fermi-liquid structure assumed for kaon condensation could be breaking down at a density corresponding to the onset of the topological phase change. An important question is then if there is a topological phase change at a density lower than a possible condensation density, how can the kaon condensation process be described from such a non-Fermi liquid state? Another question tied to the above is: Would kaon condensation, triggered by the $\Sigma$-term type interaction, induce the onset of a non-Fermi liquid state of the type discussed in condensed matter physics? If so, what happens to the EoS in that phase? We have not yet succeeded to answer these questions fully and we will come back to them in the future.

The RG analysis made here suggests that the WT term, playing an insignificant role in kaon condensation, would also be irrelevant in triggering the possible Fermi-liquid-to-non-Fermi liquid (FLNFL) transition. This follows from that the kaon is a (pseudo-)Goldstone particle and the WT term is the leading derivative coupling in kaon-Fermi-liquid interactions. And the derivative Goldstone-fermion coupling is known to prevent the FLNFL transition~\cite{no-go}.  It is the non-derivative coupling due to chiral symmetry breaking associated with the spontaneous breaking of scale symmetry that  could potentially provide a loophole to the Goldstone-boson's ``no-go theorem" for the FLNFL transition. Such loopholes are known to exist in condensed matter physics~\cite{no-go}.

A potentially promising possibility that is perhaps closely related to the role of the dilaton mentioned above is the density dependence on the coupling $h$ suggested by BR scaling. As shown by Song~\cite{song}, the density-dependence in nuclear effective Lagrangians can be made consistent with the thermodynamic consistency {\it only} if it is given as a functional of the baryon density local operator $\psi^\dagger\psi$. It seems possible in the spirit of \cite{silverstein} to have $h_{\Sigma}$ develop the scaling $[h_{\Sigma}]=-2$ via $\psi^\dagger\psi$ dependence so that the action becomes relevant like $[I_{\psi\phi}]_{h_\Sigma}$ in Eq.(\ref{II}). This is analogous to what happens in the presence of BR scaing. In chiral Lagrangians, the $\Sigma$-term goes as $\sim f_\pi^{-2}$ which increases as the pion decay constant drops at increasing density.

To conclude, we stress the basic difference between chiral perturbation theory where the expansion is made in the ``small parameter" $k_F$ whereas in the RG approach anchored on the Fermi-liquid fixed point, the expansion is in the ``small parameter" $1/k_F$. This switches the importance of the WT-like term and the $\Sigma$-like term between the two approaches. What appears to be subleading in chiral counting in ChPT is relevant -- and more important -- in the RG sense in the effective model built on the Fermi-liquid fixed point. As the result, the latter predicts that kaons could condense at a density considerably lower than in the former. If this is correct, it will be necessary that kaon condensation be taken into account in calculating the EoS in stars. The question then is: If kaons can condense at a relatively low density, what about hyperons? The other question is: How can one reconcile with the $\sim 2$-solar mass stars? These questions could be addressed in the given framework and constitutes a future project.
 \subsection*{Acknowledgments}
We are grateful for discussions and comments from Chang-Hwan Lee and Hyun Kyu Lee. One of us (MR) would like to thank Youngman Kim for arranging the visit to the theory group of  Rare Isotope Science Project of Institute for Basic Science which made this collaboration possible. The work of WGP was supported by the Rare Isotope Science Project of Institute for Basic Science funded by Ministry of Science, ICT and Future Planning and National Research Foundation of Korea (2013M7A1A1075766).
\newpage
\appendix

\centerline{\large\bf  APPENDIX}

\setcounter{section}{0}
\renewcommand{\thesection}{\Alph{section}}
\setcounter{equation}{0}
\renewcommand{\theequation}{\Alph{section}.\arabic{equation}}

\section{Scaling of the Delta Function}
\label{A}
\begin{figure}[h]
\begin{center}
\includegraphics[width=8.0cm]{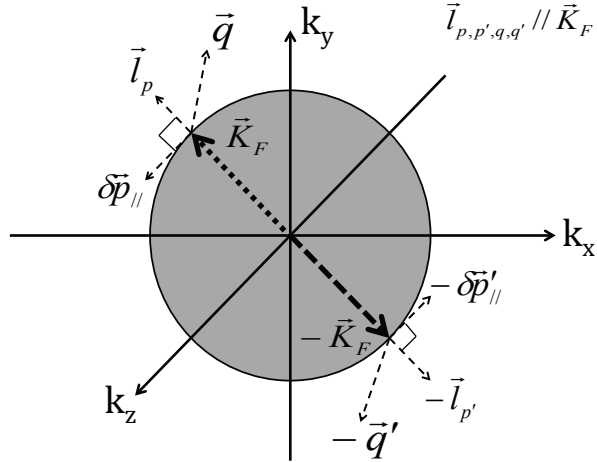}
\caption{The kinematic configuration in the momentum space.}
\label{delta_function2}
\end{center}
\end{figure}

We evaluate the scaling of the delta function following the reasoning of Polchinski~\cite{shankar}.
When the momenta of $\psi$ and $\phi$ are given as in Fig.~\ref{delta_function2}, we have the Dirac delta function in the momentum space as
\begin{equation}
\delta^{(3)}\left(\vec{p} + \vec{q} - \vec{p}^{\, \prime} - \vec{q}^{\, \prime}  \right)\,.
\end{equation}
We decompose $\vec{p}$ and $\vec{p}^{\, \prime}$ as
\begin{eqnarray}
\vec{p}
&=&
\vec{k}_F + \vec{l}_{p} + \delta\vec{p}_{\parallel}\,, \\
\vec{q}
&=&
\vec{l}_{q} + \delta\vec{q}_{\parallel}\,, \\
\vec{p}^\prime
&=&
\vec{k}_F + \vec{l}_{p^\prime} + \delta\vec{p}^{\, \prime}_{\parallel}\,, \\
\vec{q}^\prime
&=&
\vec{l}_{q^\prime} + \delta\vec{q}^{\, \prime}_{\parallel}\,,
\end{eqnarray}
the momenta can be expressed as in Fig.~\ref{delta_function2}, where $l$'s are perpendicular to the Fermi surface and
the vectors with the subscript $\parallel$ are parallel to the Fermi surface.
Note that the $l$'s are independent of the parallel components. That is,
\begin{equation}
\delta^{(3)}\left(\vec{p} + \vec{q} - \vec{p}^\prime - \vec{q}^\prime  \right) = \delta^{(2)}\left( \delta\vec{p}_{\parallel} + \delta\vec{q}_{\parallel} -\delta\vec{p}^{\, \prime}_{\parallel} - \delta\vec{q}^{\, \prime}_{\parallel}\right) \delta^{(1)}\left( \vec{l}_p + \vec{l}_{q} - \vec{l}_{p^\prime} - \vec{l}_{q^\prime} \right)\,.
\end{equation}
When the high frequency modes are integrated out up to $s\bar{\Lambda}$, the $l$'s scale as $l \rightarrow s l$ and the delta function constrains l's as
\begin{equation}
\delta^{(1)}\left( \vec{l}_p + \vec{l}_{q} - \vec{l}_{p^\prime} - \vec{l}_{q^\prime} \right) \rightarrow \delta^{(1)}\left( s\vec{l}_p +  s\vec{l}_{q} - s\vec{l}_{p^\prime} - s\vec{l}_{q^\prime} \right) = s^{-1}\delta^{(1)}\left( \vec{l}_p + \vec{l}_{q} - \vec{l}_{p^\prime} - \vec{l}_{q^\prime} \right)\,.
\end{equation}
On the other hand, for the parallel components of the delta function, we should consider two possible cases for the nucleon momentum:
\begin{enumerate}
\item
In the case 1, we have
\begin{eqnarray}
\delta^{(2)}\left( \delta\vec{p}_{\parallel} + \delta\vec{q}_{\parallel} -\delta\vec{p}^{\, \prime}_{\parallel} - \delta\vec{q}^{\, \prime}_{\parallel}\right)
&\rightarrow&
\delta^{(2)}\left( \delta\vec{p}_{\parallel} + s\delta\vec{q}_{\parallel} -\delta\vec{p}^{\, \prime}_{\parallel} - s\delta\vec{q}^{\, \prime}_{\parallel}\right)\\
&\approx& \delta^{(2)}\left( \delta\vec{p}_{\parallel} -\delta\vec{p}^{\, \prime}_{\parallel} \right)\,.
\end{eqnarray} So the delta function for the parallel components does not scale. In this case, we have
\be
[\delta^4 (p,q,p^\prime,q^\prime)]=-2.
\ee

\item
In the case 2, we have
\begin{eqnarray}
\delta^{(2)}\left( \delta\vec{p}_{\parallel} + \delta\vec{q}_{\parallel} -\delta\vec{p}^{\, \prime}_{\parallel} - \delta\vec{q}^{\, \prime}_{\parallel}\right)
&\rightarrow&
\delta^{(2)}\left(  s\delta\vec{q}_{\parallel} - s\delta\vec{q}^{\, \prime}_{\parallel}\right)\\
&=&
\left(s^{-1} \right)^2 \delta^{(2)}\left(\delta\vec{q}_{\parallel} -\delta\vec{q}^{\, \prime}_{\parallel} \right)\,,
\end{eqnarray}  where we imposed $\delta\vec{p}_{\parallel} = \delta\vec{p}^{\, \prime}_{\parallel}$.
Then,  the delta function for the parallel components scales as $\delta \rightarrow s^{-2}\delta$. In this case,
\be
[\delta^4 (p,q,p^\prime,q^\prime)]=-4.
\ee
This is the scaling of the delta function implied by the generalized hidden local symmetry approach discussed in Section \ref{hls}.
\end{enumerate}
\section{Scaling of One-Meson Exchange Diagrams}\label{one-meson}
Here we give the details of how one-meson-exchange graphs scale. We use the recipes for scaling given in Section \ref{recipe} and in \cite{kachru}. The ambiguity in scaling of the delta function $[\delta^4 (p,q,p^\prime,q^\prime)]= -2$ or $-4$ encountered above arises because it involves both quasiparticle and boson momenta that scale differently. In order to remove the ambiguity, we will avoid integrating over the delta function that contains both quasiparticle momentum and boson (both $\phi$ and $K$) momentum.

\begin{figure}[h]
\begin{center}
\includegraphics[width=12.0cm]{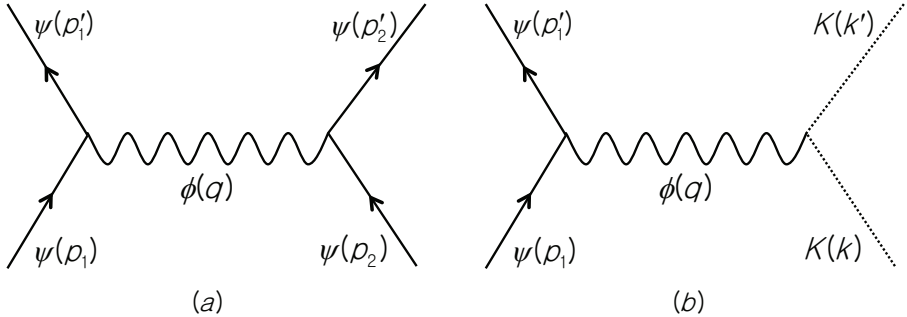}
\caption{One-$\phi$-exchange graphs for (a) the four-Fermi interaction and (b) the KN-KN interaction.}
\label{OBE}
\end{center}
\end{figure}
The graphs we consider are given in Fig.~\ref{OBE}. First look at the four-Fermi interaction Fig.~\ref{OBE}(a). Written in full detail, it has the form
\be
I_{4fermi}=&&\int (d^4p_1)(d^4p_2)(d^4p_1^\prime)(d^4p_2^\prime) (d^4q) \bar{\psi}(p_1^\prime)\bar{\psi}(p_2^\prime)\psi(p_1)\psi(p_2) \delta^{4}(p_1- p_1^\prime-q)\delta^{4}(p_2 - p_2^\prime+q)\nonumber\\
 &&\times \frac{1}{q^2 + m_\phi^2}\,. \label{4fermi}
\ee
One may be tempted to integrate over the $\phi$ momentum using the delta function but we should eschew doing this. Defining $\delta_1 \equiv \delta^{4}(p_1- p_1^\prime-q)$ and $\delta_2 \equiv \delta^{4}(p_2 - p_2^\prime+q)$, we express $I_{4fermi}$ in a compact form
\begin{equation}
I_{4fermi}=\int (d^4p)^4 (d^4q) \bar{\psi}\bar{\psi}\psi\psi \times \delta_1\delta_2 \times \frac{1}{q^2+m_\phi^2}\,.
\end{equation}
Now using the counting rules given in  Section \ref{recipe} and in  \cite{kachru} (for the $\psi\psi\phi$ vertex), we have $[d^4p]=2$, $[d^4q]=4$, $[\delta_1]=[\delta_2]=-2$, and $[\psi]=[\bar{\psi}]=-3/2$. Thus we get
\be
[I_{4fermi}]=2+[``1/(q^2+m_\phi^2)"].
\ee
In order to assure that the incorporation of $\phi$, one of the mHLS degrees of freedom, leaves the Fermi-liquid structure intact, we have to impose that $I_{4fermi}$ be marginal, i.e., $[I_{4fermi}]=0$. Therefore {\it in the form given where the $\phi$ integration is left undone}, we have the constraint
\be
[``1/(q^2+m_\phi^2)"]= -2.\label{constraint}
\ee
Note the quotation mark representing a mnemonic. One can see that this is equivalent to the scaling $[d^4q \phi^2]=-2$.

Let us now do the same analysis with the diagram of Fig.~\ref{OBE}(b).  We write the KN-KN interaction in complete parallel to the four-Fermi interaction (\ref{4fermi}),
\begin{equation}
I_{KN}=\int (d^4p)^2 (d^4k)^2 (d^4q) \bar{\psi}\psi\bar{K} K \times \delta_1\delta_K \times \frac{1}{q^2+m_\phi^2}\,.
\end{equation}
Here two quasiparticle fields on the right vertex of Fig.~\ref{OBE}(a) are replaced by two kaon fields with the corresponding integration measures and the delta function $\delta_2 \equiv \delta^{4}(p_2 - p_2^\prime+q)$ is replaced by $\delta_K \equiv \delta^{4}(k - k^\prime+q)$. Since we have $[d^4k K]=1$ and $[\delta_K]$=-4, we can immediately read off the scaling
\be
[I_{KN}]=1+[``1/(q^2+m_\phi^2)"].
\ee
 Given that the structure of the graph is identical to that of the four-Fermi interaction, we can impose the scaling (\ref{constraint}). This then gives $[I_{KN}]=-1$,  i.e., the nonlocal form of $\Sigma$-term type interaction is relevant. This discussion does not depend on what the mass of $\phi$ is, so we can take it to $\infty$, arriving at the local interaction (\ref{couple2}) in the class 2 (\ref{II}). What happens in the ``mended-symmetry limit" with $m_\phi\rightarrow 0$ is a subtle matter that needs to be treated with proper account of mHLS.

\section{Calculation of One-Loop Graph for $h_\Sigma$}\label{B}
In this appendix, we give the details of calculating the one-loop graph of Fig.~\ref{kaon_fermion} for the $\Sigma$-coupling. The details for the Weinberg-Tomozawa coupling are the same but the result will be different.

The action for the renormalization of $h$ term that we are concerned with here is given at one loop by
\begin{equation}
-\int \left(\frac{d\epsilon d^3\vec{k}}{(2\pi)^4}\right)^2 \left(\frac{d\omega d^3 \vec{q}}{(2\pi)^4} \right)^2 \, h\,\phi^*\phi \bar{\psi}\psi \delta^4(\omega,\,\epsilon,\,\vec{q},\,\vec{k}) - \delta S\,,
\end{equation} where
\begin{equation}
\delta S
=
\int \left( \frac{d\omega d^3 q}{\left(2\pi \right)^4} \right)^4 \left( \frac{d\epsilon d^3 k}{\left(2\pi \right)^4} \right)^4 h^2 \left[\phi^*\phi \bar{\psi}\psi \delta^4(\omega,\,\epsilon,\,\vec{q},\,\vec{k}) \right]^2\,.
\end{equation}
Decomposing the fields into `$<$' for low frequency mode and `$>$' for high frequency mode, and eliminating the high-frequency modes, there are four terms containing low-frequency modes contributing to $h$ which can be equated to the two graphs given in Fig.~\ref{kaon_fermion}.  For the kinematics we are interested in, the graphs (a) and (b) give the same results, so we will focus on (a). If we set the loop energy-momentum of the fermion and the meson respectively as $(E^\prime, \vec{k}^\prime)$ and $(\omega^\prime, \vec{q}^{\,\prime})$, and imposing the energy-momentum conservation
\begin{eqnarray}
E -\omega + \omega^\prime -E^\prime = 0\,, \\
\vec{k} - \vec{q} + \vec{q}^{\,\prime} - \vec{k}^{\,\prime} =0 \,,
\end{eqnarray}
we have
\begin{eqnarray}
\omega^\prime - E^\prime = \omega - E &\Rightarrow&  \omega^\prime =  \omega - E + E^\prime = \omega- \epsilon - \epsilon_F + \epsilon^\prime + \epsilon_F = \epsilon^\prime -Q_0\,, \\
\vec{k} - \vec{q}  = \vec{k}^\prime - \vec{q}^{\,\prime}  &\Rightarrow& \vec{q}^{\,\prime} =  \Omega\left(k_F + l^\prime \right) - \Omega\left(k_F + l \right)\,,
\end{eqnarray} where $Q_0 \equiv \epsilon - \omega$ and $\Omega\left(K_F + l \right) \equiv \vec{k} - \vec{q}$.
Then the loop energy-momentum is
\begin{eqnarray}
\omega^\prime &=& \epsilon^\prime -Q_0\,, \\
\vec{q}^{\,\prime} &=& \Omega\left(k_F + l^\prime \right) - \Omega\left(k_F + l \right)\,,
\end{eqnarray} where $Q_0$ is the external energy difference between the fermion and meson, satisfying the dispersion relations,
\begin{eqnarray}
\epsilon &=& \frac{\vec{k}^{\, 2}}{2m^\ast} -\epsilon_F\,, \\
\omega^2 &=& \vec{q}^{\, 2} + m_\phi^2\,.
\end{eqnarray}
From the above energy-momentum conservation, we have
\begin{equation}
\vec{q}^{\,\prime\, 2} \approx \left( l^{\prime} - l \right)^2 + 2k_F^2\left( 1 - \cos \theta \right) + {\cal O}\left( 1/k_F \right)\,.
\end{equation}
The contribution from the diagram (a) is given by
\begin{eqnarray}
\delta S^{(a)}
&\equiv&
{} \int \frac{\left(d\omega d^3q \right)^2 \left( d\epsilon d^3k \right)^2}{(2\pi)^{16}} h^2 \phi^\ast_< \phi_< \bar{\psi}_< \psi_< \, \delta^4 \left( \epsilon, \omega, \vec{k}, \vec{q}\right)
\nonumber\\
&&
{} \times \int \frac{d\epsilon^{\prime} d^3 k^{\prime}}{(2\pi)^4} \frac{-1}{i\epsilon^{\prime} - \vec{v}_F \cdot \vec{l}^\prime} \, \frac{1}{ \left(\epsilon^\prime - Q_0\right)^2 + \left( l^{\prime} - l \right)^2 + 2k_F^2\left( 1 - \cos \theta \right) + m_\phi^2 }\,,
\\
&\approx&
{} \int \frac{\left(d\omega d^3q \right)^2 \left( d\epsilon d^3k \right)^2}{(2\pi)^{16}} h^2 \phi^\ast_< \phi_< \bar{\psi}_< \psi_< \, \delta^4 \left( \epsilon, \omega, \vec{k}, \vec{q}\right)
\nonumber\\
&&
{} \times \int \frac{d\epsilon^{\prime}\, k_F^2\, d l^{\prime} d^2 \Omega}{(2\pi)^4} \frac{-1}{i\epsilon^{\prime} - \vec{v}_F \cdot \vec{l}^\prime} \, \frac{1}{ \left(\epsilon^\prime - Q_0\right)^2 +\left( l^{\prime} - l \right)^2 + m_\phi^{\prime\, 2} }\,,
\end{eqnarray} where we used $k^{\prime\, 2} = \left( \vec{k}_F + \vec{l}^\prime \right)^2 \approx k_F^2$ with $\bar{\Lambda} \ll k_F$ and defined $m_\phi^{\prime\, 2} \equiv m_\phi^2 + 2k_F^2\left( 1 - \cos \theta \right)$.
Let us evaluate the integral
\begin{eqnarray}
{\cal I}
&\equiv&
{} \int \frac{d\epsilon^{\prime}\, k_F^2\, d l^{\prime} d^2 \Omega}{(2\pi)^4} \frac{1}{ \left(\epsilon^\prime - Q_0\right)^2 +\left( l^{\prime} - l \right)^2 + m_\phi^{\prime\, 2} }\,  \frac{-1}{i\epsilon^{\prime} - \vec{v}_F \cdot \vec{l}^\prime } \\
&=&
{} \frac{ k_F^2}{(2\pi)^4}\int^\infty_{-\infty} d\epsilon^{\prime}\, \int_{s\bar{\Lambda} < \left|l^\prime \right| < \bar{\Lambda}}  \frac{d^2 \Omega \,d l^{\prime} }{ \left(\epsilon^\prime - Q_0\right)^2 +\left( l^{\prime} - l \right)^2 + m_\phi^{\prime\, 2} }\, \frac{-1}{i\epsilon^{\prime} - \vec{v}_F \cdot \vec{l}^\prime }\,.\label{Iinteg}
\end{eqnarray}
Since the RGE of $h$ should be independent of $Q_0$ and $l$,  we set $Q_0 = l = 0$ in Eq.~(\ref{Iinteg}).
Integrating over $\epsilon^\prime$, we get
\begin{eqnarray}
{\cal I}_{\pm} &=& \frac{k_F^2}{\left(2\pi \right)^3} \int d^2 \Omega \int_{s\bar{\Lambda} < \left|l^\prime \right| < \bar{\Lambda}} dl^\prime \nonumber \\
         && \times (-1)\left[ \frac{ \pm\theta\left( \mp v_F l^\prime  \right) \mp \frac{1}{2}}{ \left( 1 - v_F^2 \right)l^{\prime\, 2} + m_\phi^{\prime\, 2}} + \frac{\frac{1}{2} v_F l^\prime}{ \left( 1 - v_F^2\right) l^{\prime\, 2} + m_\phi^{\prime\, 2} } \frac{1}{ \sqrt{l^{\prime \, 2} + m_\phi^{\prime\, 2}}} \right]\,,
\end{eqnarray} where ${\cal I}_\pm$ is for the upper/lower hemisphere in integrating $\epsilon^\prime$ in the $\epsilon^\prime$ complex plane.

In doing the integration over $l^\prime$, we use the relations\footnote{These relations are valid for any $f(l^\prime)$. If we define a function of $l^\prime$, $F(l^\prime)$, as
\begin{equation}
\frac{d\, F(l^\prime)}{dl^\prime} \equiv f(l^\prime)\,,
\end{equation}
\begin{eqnarray}
\int^{\bar{\Lambda}}_{\bar{\Lambda}(1-\delta t)} dl^\prime f(l^\prime)
&=&
\frac{F\left( \bar{\Lambda} \right) - F\left( \bar{\Lambda} -\bar{\Lambda} \delta t  \right)}{\bar{\Lambda} \delta t} \bar{\Lambda} \delta t \\
&=&
f\left( \bar{\Lambda} \right) \bar{\Lambda} \delta t
\end{eqnarray} in the limit of $\delta t \rightarrow 0$.  },
\begin{eqnarray}
\int_{s\bar{\Lambda} < \left|l^\prime \right| < \bar{\Lambda}} d l^\prime f(l^\prime) &=& \int^{\bar{\Lambda}}_{\bar{\Lambda}(1-\delta t)} dl^\prime f(l^\prime) + \int^{-\bar{\Lambda}(1-\delta t)}_{-\bar{\Lambda}} dl^\prime f(l^\prime)\,, \\
\int^{\bar{\Lambda}}_{\bar{\Lambda}(1-\delta t )} dl^\prime f(l^\prime)&=& f(\bar{\Lambda}) \bar{\Lambda} \delta t\,,\\
\int^{-\bar{\Lambda}(1-\delta t)}_{-\bar{\Lambda}} dl^\prime f(l^\prime)&=& f(-\bar{\Lambda}) \bar{\Lambda} \delta t \,.
\end{eqnarray}
After integrating over $l^\prime$, we get
\begin{eqnarray}
{\cal I}_\pm &=& \left( - \bar{\Lambda} \delta t  \right) k_F^2 \int \frac{d^2 \Omega}{(2\pi)^3}\nonumber\\
             &\times& \left[ \frac{ \pm\theta\left( \mp v_F \bar{\Lambda}  \right) \mp \frac{1}{2}}{ \left( 1 - v_F^2 \right)\bar{\Lambda}^2 + m_\phi^{\prime\, 2}} + \frac{\frac{1}{2} v_F \bar{\Lambda}}{ \left( 1 - v_F^2\right) \bar{\Lambda}^2 + m_\phi^{\prime\, 2} } \frac{1}{ \sqrt{\bar{\Lambda}^2 + m_\phi^{\prime\, 2}}} \right. \nonumber \\
             &&\left.+ \frac{ \pm\theta\left( \pm v_F \bar{\Lambda}  \right) \mp \frac{1}{2}}{ \left( 1 - v_F^2 \right)\bar{\Lambda}^2 + m_\phi^{\prime\, 2}} - \frac{\frac{1}{2} v_F \bar{\Lambda}}{ \left( 1 - v_F^2\right) \bar{\Lambda}^2 + m_\phi^{\prime\, 2} } \frac{1}{ \sqrt{\bar{\Lambda}^2 + m_\phi^{\prime\, 2}}}  \right]\\
             &=& 0\,.
\end{eqnarray}

\end{document}